\documentclass{mi_aa}
\usepackage{graphicx}
\usepackage{times,float}

\begin{document}

\title{The bright optical afterglow of the long \object{GRB 001007}\thanks{Based on observations collected: at the European Southern Observatory, in La Silla and Paranal (Chile), ESO Large Programmes 165.H-0464(G), 165.H-0464(I) and 265.D-5742(A), granted to the GRACE Team; with the Livermore Optical Transient Imaging System's 0.11 m telephoto lenses, at the Lawrence Livermore National Laboratory, in California (USA); with the Instituto de Astrof\'\i sica de Canarias's 0.82 m telescope, at the Observatorio del Teide, in the island of Tenerife (Spain); with the Danish 1.54 m telescope, at the European Southern Observatory, in La Silla (Chile); and with the Nordic Optical Telescope, operated on the island of La Palma jointly by Denmark, Finland, Iceland, Norway, and Sweden, in the Spanish Observatorio del Roque de los Muchachos, of the Instituto de Astrof\'\i sica de Canarias.}}

\author{J.M.     Castro Cer\'on       \inst{1}
   \and A.J.     Castro-Tirado        \inst{2}
   \and J.       Gorosabel            \inst{2,3,4}
   \and J.       Hjorth               \inst{5}
   \and J.U.     Fynbo                \inst{6}
   \and B.L.     Jensen               \inst{5}
   \and H.       Pedersen             \inst{5}
   \and     M.I.~Andersen             \inst{7}
   \and M.       L\'opez-Corredoira   \inst{8}
   \and O.       Su\'arez             \inst{4,9}
   \and Y.       Grosdidier           \inst{8}
   \and J.       Casares              \inst{8}
   \and D.       P\'erez-Ram\'{\i}rez \inst{10}
   \and       B.~Milvang-Jensen       \inst{11}
   \and G.       Mall\'en-Ornelas     \inst{12}
   \and A.       Fruchter             \inst{13}
   \and J.       Greiner              \inst{14}
   \and E.       Pian                 \inst{15}
   \and P.M.     Vreeswijk            \inst{16}
   \and     S.D.~Barthelmy            \inst{17}
   \and T.       Cline                \inst{17}
   \and F.       Frontera             \inst{18}
   \and L.       Kaper                \inst{16}
   \and S.       Klose                \inst{19}
   \and C.       Kouveliotou          \inst{20}
   \and D.H.     Hartmann             \inst{21}
   \and       K.~Hurley               \inst{22}
   \and N.       Masetti              \inst{18}
   \and E.       Mazets               \inst{23}
   \and E.       Palazzi              \inst{18}
   \and H.S.     Park                 \inst{24}
   \and E.       Rol                  \inst{16}
   \and I.       Salamanca            \inst{16}
   \and N.       Tanvir               \inst{25}
   \and     J.I.~Trombka              \inst{17}
   \and R.A.M.J. Wijers               \inst{26}
   \and G.G.     Williams             \inst{27}
   \and E.       van den Heuvel       \inst{16}
   }

\offprints{J.M. Castro Cer\'on, \\
    \email{josemari@alumni.nd.edu}
    }

\institute{Real Instituto y Observatorio de la Armada, Secci\'on de Astronom\'\i a, 11.110 San Fernando-Naval (C\'adiz) Spain.
           \email{josemari@alumni.nd.edu}
      \and Instituto de Astrof\'\i sica de Andaluc\'\i a (CSIC), Apartado de Correos, 3.004, 18.080 Granada Spain.
           \email{ajct@iaa.es,jgu@iaa.es}
      \and Danish Space Research Institute, Juliane Maries Vej 30, 2\,100 Copenhagen \O\ Denmark.
           \email{jgu@dsri.dk}
      \and Laboratorio de Astrof\'\i sica Espacial y F\'\i sica Fundamental (INTA), Apartado de Correos, 50.727, 28.080 Madrid Spain.
           \email{jgu@laeff.esa.es}
      \and Astronomical Observatory, University of Copenhagen, Juliane Maries Vej 30, 2\,100 Copenhagen \O\ Denmark. \\
           \email{jens@astro.ku.dk,brian\_j@astro.ku.dk,holger@astro.ku.dk}
      \and European Southern Observatory, Karl Schwarzschild Stra\ss e 2, 85\,748 Garching Germany.
           \email{jfynbo@eso.org}
      \and Division of Astronomy, PO Box 3\,000, 90\,014 University of Oulu Finland.
           \email{michael.andersen@oulu.fi}
      \and Instituto de Astrof\'\i sica de Canarias, 38.200 La Laguna (Tenerife) Spain.
           \email{martinlc@ll.iac.es,jcv@ll.iac.es,yves@ll.iac.es}
      \and Departamento de Ciencias de la Navegaci\'on y de la Tierra, Universidad de la Coru\~na, 15.011 La Coru\~na Spain.
           \email{olga@mail2.udc.es}
      \and Department of Physics, Michigan Technological University, 1\,400 Townsend Drive, Houghton MI 49\,931-1\,295 USA.
           \email{dperez@mtu.edu}
      \and School of Physics and Astronomy, University of Nottingham, University Park NG7 2RD Nottingham UK.
           \email{milvang@astro.ku.dk}
      \and Department of Astronomy, University of Toronto, 60 St. George Street, Toronto ON M5S 3H8 Canada.
           \email{mallen@astro.utoronto.ca}
      \and Space Telescope Science Institute, 3\,700 San Mart\'\i n Drive, Baltimore MD 21\,218 USA.
           \email{fruchter@stsci.edu}
      \and Astrophysikalisches Institut, An der Sternwarte 16, 14\,482 Potsdam Germany.
           \email{jgreiner@aip.de}
      \and Osservatorio Astronomico di Trieste, Via Tiepolo 11, 34\,131 Trieste Italy.
           \email{pian@tesre.bo.cnr.it}
      \and University of Amsterdam, Kruislaan 403, 1\,098 SJ Amsterdam The Netherlands. \\
           \email{pmv@astro.uva.nl,lexk@astro.uva.nl,evert@astro.uva.nl,isabel@astro.uva.nl,edvdh@astro.uva.nl}
      \and Goddard Space Flight Centre (NASA), Greenbelt MD 20\,771 USA. \\
           \email{scott@milkyway.gsfc.nasa.gov,thomas.cline@gsfc.nasa.gov,jack.trombka@gsfc.nasa.gov}
      \and Istituto di Astrofisica Spaziale e Fisica Cosmica, Sezione di Bologna (CNR), Via Gobetti 101, 40\,129 Bologna Italy. \\
           \email{filippo@tesre.bo.cnr.it,masetti@tesre.bo.cnr.it,eliana@tesre.bo.cnr.it}
      \and Th\"uringer Landessternwarte Tautenburg, 07\,778 Tautenburg Germany.
           \email{klose@tls-tautenburg.de}
      \and Marshall Space Flight Centre (NASA), SD-50, Huntsville AL 35\,812 USA.
           \email{kouveliotou@eagles.msfc.nasa.gov}
      \and Clemson University, Clemson SC 29\,634 USA.
           \email{hartmann@grb.phys.clemson.edu}
      \and University of California, Berkeley, Space Sciences Laboratory, Berkeley CA 94\,720-7\,450 USA.
           \email{khurley@ssl.berkeley.edu}
      \and Ioffe Physico-Technical Institute, 26 Polytekhnicheskaya, St. Petersburg 194\,021 Russia.
           \email{mazets@astro.ioffe.rssi.ru}
      \and Lawrence Livermore National Laboratory, Livermore CA 94\,550 USA.
           \email{park@ursa.llnl.gov}
      \and Department of Physical Sciences, University of Hertfordshire, College Lane, Hatfield, Herts AL10 9AB UK.
           \email{nrt@star.herts.ac.uk}
      \and Department of Physics and Astronomy, State University of New York, Stony Brook NY 11\,794-3\,800 USA.
           \email{rwijers@astro.sunysb.edu}
      \and Steward Observatory, Tucson AZ 85\,721 USA.
           \email{gwilliams@as.arizona.edu}
           }

\date{Received 25 March 2002 / Accepted 8 July 2002}

\abstract{We present optical follow up observations of the long \object{GRB 001007} between 6.14 hours and $\sim$ 468 days after the event. An unusually bright optical afterglow (OA) was seen to decline following a steep power law decay with index $\alpha$ = $-$2.03 $\pm$ 0.11, possibly indicating a break in the light curve at $t - t_\mathrm{0} <$ 3.5 days, as found in other bursts. Upper limits imposed by the LOTIS alerting system 6.14 hours after the gamma ray event provide tentative (1.2$\sigma$) evidence for a break in the optical light curve. The spectral index $\beta$ of the OA yields $-$1.24 $\pm$ 0.57. These values may be explained both by several fireball jet models and by the cannonball model. Fireball spherical expansion models are not favoured. Late epoch deep imaging revealed the presence of a complex host galaxy system, composed of at least two objects located 1.2\arcsec ~(1.7$\sigma$) and 1.9\arcsec ~(2.7$\sigma$) from the afterglow position.
         }

\maketitle

\section{Introduction}

Gamma Ray Bursts (GRBs hereafter) are flashes of high energy ($\sim$ 1 keV--10 GeV) photons (\cite{fishman-meegan95}), occurring at cosmological distances. Observations of GRBs are well described by the so called fireball model in which an explosive flow of relativistic ejecta is released from a central unknown source. Shortly thereafter, the fast moving ejecta sweeps the interstellar matter. The shocked gas powers long lived, broad band emission: the afterglow. This is primarily synchrotron radiation. According to the current view, a forward external shock wave would have produced the afterglow as observed at all wavelengths. So far, $\sim$ 3\,000 GRBs have been detected in $\gamma$~rays. Of these $\sim$ 100 have been localised to arcminute accuracy and deeply followed up; but only $\sim$ 25 have been pinpointed at optical wavelengths, with redshifts ranging from $z$ = 0.0085 (\cite{galama98}) to $z$ = 4.50 (\cite{andersen00}). The population of electrons is assumed to be a power law distribution of Lorentz factors $\Gamma_e$, following d$N$/d$\Gamma_\mathrm{e} \propto \Gamma_\mathrm{e}^{-p}$ above a minimum Lorentz factor $\Gamma_\mathrm{e} \geq \Gamma_m$, corresponding to the synchrotron frequency $\nu_m$. The electron distribution power law index is given by $p$, ranging from 2 to 2.5 for afterglows observed to date (\cite{van-paradijs00}). The electrons radiate synchrotron emission with a flux density $F_\nu \propto (t - t_\mathrm{0})^\alpha \nu^\beta$, where $\alpha$ is the temporal decay index, $\beta$ is the spectral index and $(t - t_\mathrm{0})$ represents the time elapsed since the GRB onset in gamma rays.

\begin{figure}
      \resizebox{\hsize}{!}{\includegraphics{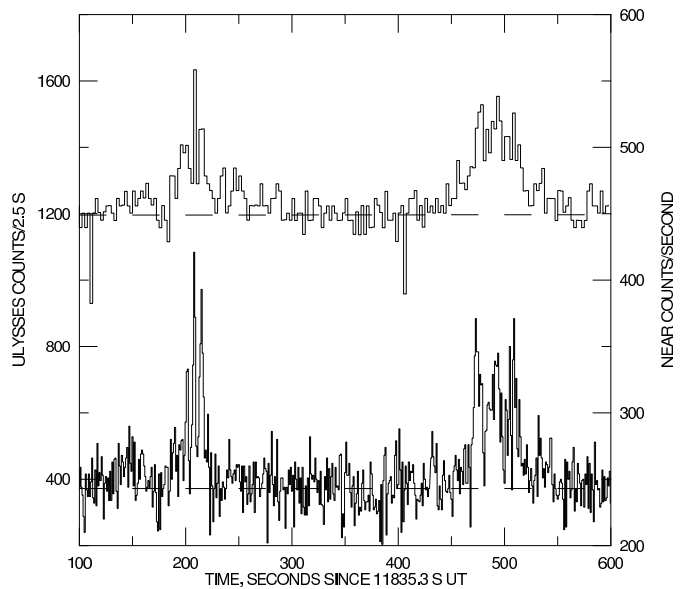}}
                 \caption{Ulysses (upper plot) and NEAR (lower plot) time histories for \object{GRB 001007}. The light curves are accumulated between 25--150, and $>$150 keV for Ulysses and NEAR, respectively. The burst consisted of two peaks separated by about 300 s. Dashed lines indicate the backgrounds.}
      \label{secuencia temporal}
  \end{figure}

\object{GRB 001007} was detected on 7.207488 UT October 2000 ($t_\mathrm{0}$ hereafter) by the Interplanetary Network (IPN), composed of the Ulysses, Konus-Wind and NEAR spacecraft (\cite{hurley00}). It exhibited a duration of $\sim$ 375~s, a fluence (25--100 keV) of $\sim$ 3.3$\times$10$^{-5}$ erg cm$^{-2}$ and a peak flux over 0.5 s of 7.1$\times$10$^{-7}$~erg~cm$^{-2}$~s$^{-1}$. The time history of the GRB as seen by Ulysses and NEAR is presented in Fig. \ref{secuencia temporal}. An optical afterglow (OA) was first detected by \cite{price00a}, then confirmed with further observations, in optical wavelengths, by \cite{castro-tirado00} and \cite{price00b} and, in radio wavelengths, by \cite{frail-berger00}. Section \ref{observaciones} describes the optical follow up imaging of the \object{GRB 001007} IPN error box. In Sect. \ref{resultados} we show the characteristics of the OA and its potential host galaxy. The final conclusions are listed in Sect. \ref{conclusiones}.

\section{Observations}
 \label{observaciones}

Prompt follow up observations started at $t_\mathrm{0}$ + 6.14 hours since the GRB onset, with the 0.11 m telephoto lenses of the Livermore Optical Transient Imaging System (0.11LOTIS), in California (USA). 0.11LOTIS uses a 2 $\times$ 2 CCD array. Each CCD is a 2\,048 $\times$ 2\,048 Loral 442A, giving a 8.8\degr $\times$ 8.8\degr~individual field of view (FOV). Collectively the four CCDs provide a 17.4\degr $\times$ 17.4\degr ~FOV.

Target of Opportunity programmes were triggered starting at $t_\mathrm{0}$ + 93.26 hours at, the 0.82 m telescope of the Instituto de Astrof\'\i sica de Canarias (0.82IAC) at the Observatorio del Teide, in the island of Tenerife (Spain), the 1.54 m Danish telescope (1.54D) at the European Southern Observatory, in La Silla (Chile) and the 2.56 m Nordic Optical Telescope (2.56NOT) at the Observatorio del Roque de los Muchachos, in the island of La Palma (Spain). The CCD used at the 0.82IAC is a 1\,024 $\times$ 1\,024 Thomson, giving a 7.3\arcmin $\times$ 7.3\arcmin ~FOV. The CCD used at the 1.54D (+DFOSC) is a 2\,048 $\times$ 2\,048 MAT/EEV, giving a 13.8\arcmin $\times$ 13.8\arcmin ~FOV. The CCD used at the 2.56NOT (+ALFOSC) is a 2\,048 $\times$ 2\,048 Loral/Lesser, giving a 6.5\arcmin~$\times$~6.5\arcmin ~FOV.

Further deep images were acquired in order to detect the underlying host galaxy with the European Southern Observatory 3.60 m telescope (3.60ESO), in La Silla (Chile). The CCD used at the 3.60ESO (+EFOSC2) is a 2\,048 $\times$ 2\,048 Loral, giving a 5.4\arcmin $\times$ 5.4\arcmin ~FOV. More recently, deep $R$ band images have been obtained at the 8.20 m UT3 of the Very Large Telescope (8.20VLT), 468 days after the burst. The CCD used at the 8.20VLT (+FORS1) is a 2\,048 $\times$ 2\,048 TK2048EB4-1, giving a 6.8\arcmin $\times$ 6.8\arcmin ~FOV. Table \ref{tabla1} displays the observing log.

\begin{table*}
      \begin{center}
            \caption{Journal of observations of the \object{GRB 001007} field.}
      \begin{tabular}{@{}lccccc@{}}

Date UT                  & Telescope          &Filtre&Seeing&Exposure Time     & Magnitude                                  \\
                         &                    &     &(arcseconds)&(seconds)    &                                            \\

\hline
07.4632--07.4645/10/2000 & 0.11LOTIS (CCD)    & $V$ & $\S$ & 2 $\times$     50 &                  $>$ 15.52 $^\star$        \\
08.4630--08.4643/10/2000 & 0.11LOTIS (CCD)    & $V$ & $\S$ & 2 $\times$     50 &                  $>$ 15.52 $^\star$        \\
11.0932--11.2222/10/2000 & 0.82IAC   (CCD)    & $R$ & 2.50 & 4 $\times$ 1\,800 & 20.50 $\pm$ 0.13                           \\
11.1151--11.2008/10/2000 & 0.82IAC   (CCD)    & $B$ & 2.65 & 2 $\times$ 1\,800 & 21.39 $\pm$ 0.20                           \\
11.2232--11.2440/10/2000 & 0.82IAC   (CCD)    & $V$ & 2.75 &            1\,800 & 20.96 $\pm$ 0.25                           \\
11.3192--11.3673/10/2000 & 1.54D     (DFOSC)  & $R$ & 0.95 & 4 $\times$    600 & 20.77 $\pm$ 0.06                           \\
12.2517--12.2641/10/2000 & 0.82IAC   (CCD)    & $R$ & 2.85 &            1\,069 &                  $>$ 20.30 $^{\star\star}$ \\
20.2366--20.2437/10/2000 & 2.56NOT   (ALFOSC) & $R$ & 1.00 &               600 &                  $>$ 22.10 $^{\star\star}$ \\
29.2584--29.3074/10/2000 & 1.54D     (DFOSC)  & $R$ & 0.85 & 6 $\times$    600 & 23.98 $\pm$ 0.17                           \\
02.1368--02.1676/11/2000 & 3.60ESO   (EFOSC2) & $R$ & 0.70 & 5 $\times$    500 & 24.00 $\pm$ 0.16                           \\
02.1683--02.2049/11/2000 & 3.60ESO   (EFOSC2) & $B$ & 0.90 & 5 $\times$    600 & 24.90 $\pm$ 0.25                           \\
09.0788--09.0892/11/2000 & 0.82IAC   (CCD)    & $V$ & 2.25 &               900 &                  $>$ 21.25 $^{\star\star}$ \\
16.1203--16.1718/11/2000 & 3.60ESO   (EFOSC2) & $R$ & 0.60 & 7 $\times$    600 & 24.53 $\pm$ 0.22                           \\
13.3770--13.4092/09/2001 & 3.60ESO   (EFOSC2) & $B$ & 0.85 & 3 $\times$    900 &                            $^\dag$         \\
14.2787--14.2999/09/2001 & 3.60ESO   (EFOSC2) & $B$ & 0.70 & 2 $\times$    900 & 25.10 $\pm$ 0.25                           \\
18.0544--18.0867/01/2002 & 8.20VLT   (FORS1)  & $R$ & 0.65 & 8 $\times$    300 & 24.84 $\pm$ 0.15                           \\
\hline

\multicolumn{6}{l}{$\S$            Irrelevant given the pixel scale, 15 arcseconds/pixel.}                                                                         \\
\multicolumn{6}{l}{$^\star$        10$\sigma$ upper limit (3$\sigma$ upper limits can not be quoted because 0.11LOTIS undersamples the point spread function}      \\
\multicolumn{6}{l}{                due to its large pixel scale. The 3$\sigma$ limit would not be much different than a single pixel noise variation).}            \\
\multicolumn{6}{l}{$^{\star\star}$ 3$\sigma$ upper limit.}                                                                                                         \\
\multicolumn{6}{l}{$^\dag$         The images from 13--14/09/2001 were coadded because of the faintness of the host galaxy, resulting in just a sin-}              \\
\multicolumn{6}{l}{                gle magnitude, $B$ = 25.10 $\pm$ 0.25.}                                                                                         \\

\hline

          \label{tabla1}
      \end{tabular}
      \end{center}
\end{table*}

\begin{figure}
      \resizebox{\hsize}{!}{\includegraphics{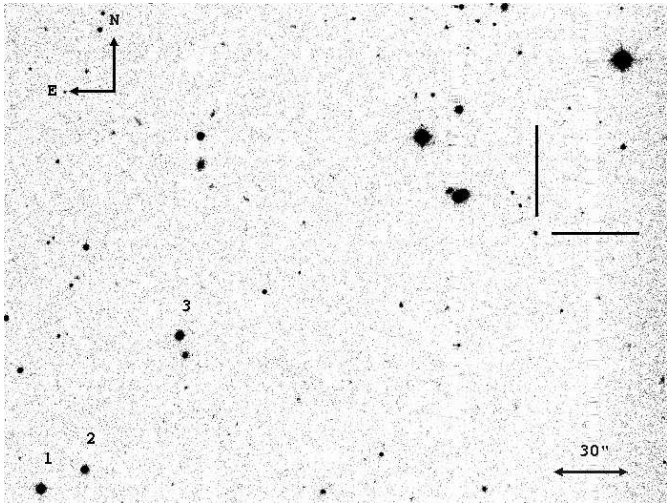}}
                 \caption{$R$ band image of the \object{GRB 001007} field taken at the 1.54D on 11.3192--11.3673 October 2000. The position of the OA is indicated between tick marks. The numbered stars represent secondary standards logged in Table \ref{tabla2}. North is upwards and East leftwards.}
      \label{estrellas secundarias}
  \end{figure}

We performed aperture photometry using the PHOT routine under IRAF\footnote{IRAF is distributed by the National Optical Astronomy Observatories, which are operated by the Association of Universities for Research in Astronomy, Inc., under cooperative agreement with the US National Science Foundation.}. The field was calibrated observing the Landolt fields \object{Rubin 149} and \object{T Phe} (\cite{landolt92}) at airmasses similar to that of the GRB field. Table \ref{tabla2} shows the position and magnitude of three secondary standards close to the OA position. Their positions on the sky are indicated in Fig. \ref{estrellas secundarias}.

\begin{table*}
      \begin{center}
            \caption{Secondary standards in the field of \object{GRB 001007}.}
      \begin{tabular}{@{}lccccc@{}}

  &  RA(J2000)  &  Dec(J2000)   & $B$              & $V$              & $R$             \\
  &  h~~~m~~~s  &\degr~~~\arcmin~~~\arcsec&        &                  &                 \\

\hline

1 & 04 06 09.26 & $-$21 55 23.9 & 19.49 $\pm$ 0.07 & 18.30 $\pm$ 0.05 & 17.19 $\pm$ 0.02 \\
2 & 04 06 07.94 & $-$21 55 16.6 & 20.02 $\pm$ 0.07 & 18.88 $\pm$ 0.08 & 18.05 $\pm$ 0.03 \\
3 & 04 06 04.98 & $-$21 54 22.5 & 19.79 $\pm$ 0.07 & 18.49 $\pm$ 0.05 & 17.44 $\pm$ 0.03 \\

\hline

          \label{tabla2}
      \end{tabular}
      \end{center}
\end{table*}

\section{Results and discussion}
  \label{resultados}

A bright OA ($R$ = 20.50, 3.95 days after the GRB, bright with respect to other OAs) was detected in the first 0.82IAC images at the preliminary position given by \cite{price00a}. An astrometric solution based on 50 USNO A2-0 reference stars in the 1.54D image taken on 11.3192--11.3673 UT October 2000 yields for the OA $\alpha_\mathrm{2000}$ = 4$^\mathrm{h}$5$^\mathrm{m}$54.28$^\mathrm{s}$, $\delta_\mathrm{2000}$ = $-$21\degr 53\arcmin 45.4\arcsec. The internal error of the position is 0.55\arcsec, which has to be added to the $1\sigma$ systematic error of the USNO catalogue ($\simeq$~0.25\arcsec according to \cite{assafin01} and \cite{deutsch99}). The final astrometric error corresponds to 0.60\arcsec. The upper limits and optical magnitudes of the afterglow are displayed in the far right column of Table \ref{tabla1}.

\subsection{The spectral shape of the OA: determination of the electron distribution power law index}
     \label{forma espectral}

We have determined the flux density distribution of \object{GRB 001007} OA on 11.15 UT October 2000 (mean epoch of the first $B$ band image) by means of our $BVR$ broad band photometric measurements obtained with the 0.82IAC. We fitted the observed flux density distribution with a power law $F_\nu \propto \nu^\beta$, where $F_\nu$ is the flux density at frequency $\nu$, and $\beta$ is the spectral index. The optical flux at the wavelengths of $B$, $V$ and $R$ bands has been derived by subtracting the contribution of the host galaxy (see Sect. \ref{presunta galaxia anfitriona}), assuming a reddening E($B - V$) = 0.042 from the DIRBE/IRAS dust maps (\cite{schlegel98}) and, correcting for the epoch differences (assuming $\alpha$ = $-$2.03 $\pm$ 0.11, calculated in Sect. \ref{curva de luz}). In converting the magnitude into flux density, the effective wavelengths and normalisations given in \cite{fukugita95} were used (assuming $V$ = 0.03, $B - V$ = 0, $V - R_\mathrm{c}$ = 0 for $\alpha$ Lyr). The flux densities are 13.0, 17.5, and 21.3 $\mu$Jy at the $B$, $V$ and $R$ bands, corrected by Galactic reddening (but not for possible intrinsic absorption in the host galaxy). The fit to the optical data $F_\nu \propto \nu^\beta$ gives $\beta$ = $-$1.24 $\pm$ 0.57. Strictly speaking, the derived value of $\beta$ is a lower limit to the actual value since we have not considered the effect of intrinsic absorption.

In the expression of the spectral index ($F_\nu \propto \nu^\beta$) $\beta$ only depends on $p$ and is independent of the geometry of the expansion. For an adiabatic expansion it is given by $\beta$ = (1 $- p$)/2 for $\nu_c > \nu > \nu_m$ and by $\beta$ = $-p$/2 for $\nu > \nu_c$, where $\nu$ is the observing frequency and $\nu_c$ is the cooling break frequency (\cite{sari98}).

The values we have derived for $p$ are 3.48 $\pm$ 1.14 ($\nu_c > \nu > \nu_m$), and 2.48 $\pm$ 1.14 ($\nu > \nu_c$). The value of $p$ for the afterglows detected to date range from 2 to 2.5 (\cite{van-paradijs00}). Thus, given the values of $p$ derived above, we consider that our $BVR$ band frequencies better agree with the $\nu > \nu_c$ regime.

\subsection{The light curve of \object{GRB 001007} OA}
     \label{curva de luz}

\begin{figure}
      \resizebox{\hsize}{!}{\includegraphics{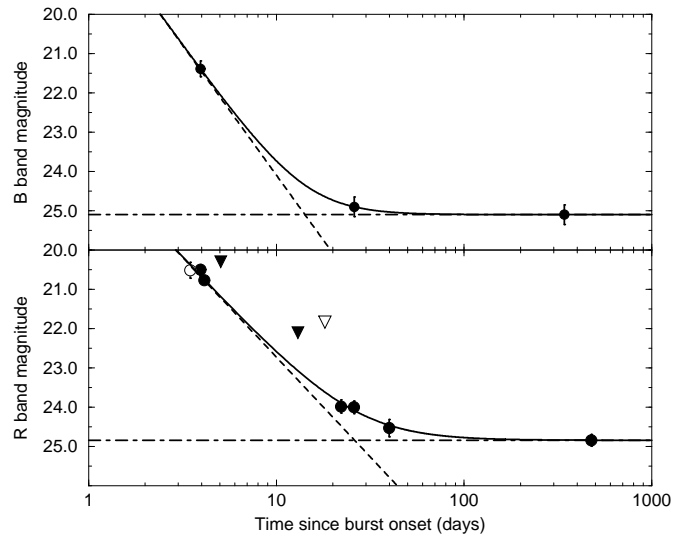}}
                 \caption{$B$ and $R$ band light curves of the OA related to \object{GRB001007}, including the underlying galaxy. Circles represent measured magnitudes and triangles represent upper limits. Filled symbols are our data points whereas empty ones are data points from the literature, the circle is from \cite{price00a} and the triangle from \cite{price00b}. These magnitudes reported by Price have been dimmed to our photometric zero point ($\Delta$m = 0.23). The short dashed and dot-dashed lines show the two components of our fit, the OA component (with $\alpha_B$ = $-$2.65 $\pm$ 0.73 and $\alpha_R$ = $-$2.03 $\pm$ 0.11) and the host galaxy component (with $B$ = 25.10 and $R$= 24.84). The sum of both components yields the solid line.}
      \label{curvas de luz}
  \end{figure}

Our $B$ and $R$ band light curves (Fig. \ref{curvas de luz}) show that the source was declining in brightness. The optical decay slowed down in early November 2000 suggesting the presence of an underlying source of constant brightness: the potential host galaxy. The deep 8.20VLT images taken in January 2002 confirmed this fact.

Most GRB optical counterparts appear to be well characterised by a power law decay plus a constant flux component, $F(t) \propto (t - t_\mathrm{0})^\alpha +F_\mathrm{host}$ where, $F(t)$ is the total measured flux density of the counterpart at time ($t - t_\mathrm{0}$) after the onset of the event at $t_\mathrm{0}$, $\alpha$ is the temporal decay index and $F_\mathrm{host}$ is the flux density of the underlying host galaxy. $F_\mathrm{host}$ can be fixed or considered as a free parametre.

In order to test the self consistency of our $R$ band data, we considered all the $R$ band points except that late VLT observation leaving $F_\mathrm{host}$ as a free parametre. By means of fitting least squares linear regression to the observed $R$ band fluxes, we predict for the host galaxy $R = 24.93 \pm 0.22$. This value is fully consistent with the $R$ band magnitude measured with the VLT ($R$ = 24.84 $\pm$ 0.15), supporting the VLT identification as the host galaxy. Fixing the $R$ band host galaxy magnitude to $R = 24.84 \pm 0.15$, a power law decline with $\alpha_R$ = $-$2.03 $\pm$ 0.11 ($\chi^2$/dof = 0.64) provides a good fit to the $R$ band data (see lower panel of Fig. \ref{curvas de luz}).

This value is also consistent with the $B$ band observations, which can be fitted by a power law decline with $\alpha_B$ = $-$2.65 $\pm$ 0.73 (the host magnitude has been fixed at $B$ = 25.10 $\pm$ 0.25; based on our late epoch imaging, see Sect. \ref{presunta galaxia anfitriona}). Thus, our data are consistent with an achromatic decay.

In order to see if a break exists in the light curve we have made use of the constraints derived from the early data taken with 0.11LOTIS. We have extrapolated the $R$~band light curve to the epoch of the first 0.11LOTIS observations (7.4638~UT October 2000). The prediction yields $R$=14.63$\pm$0.34. This value has been dereddened from Galactic extinction (\cite{schlegel98}) and then extrapolated to the $V$~band, making use of the derived value of $\beta$ (see Sect. \ref{forma espectral}). Finally, this value is reddened back in order to make it comparable to the 0.11LOTIS $V$ band upper limit measurement. This procedure yielded $V$ = 15.09 $\pm$ 0.36. This value is {\bf $1.2\sigma$} above the upper limit imposed by 0.11LOTIS (see Table \ref{tabla1}). Therefore, the 0.11LOTIS upper limit is compatible with the existence of a break in the light curve.

If the contribution of an underlying supernova were to be present in the light curve then, it is expected to peak at $\sim$~15(1+$z$) days. \object{GRB 001007} is a good candidate for such a search due to its rapid decay but, the sparse $R$ band coverage makes this search elusive.

\begin{figure*}
      \resizebox{\hsize}{!}{\includegraphics{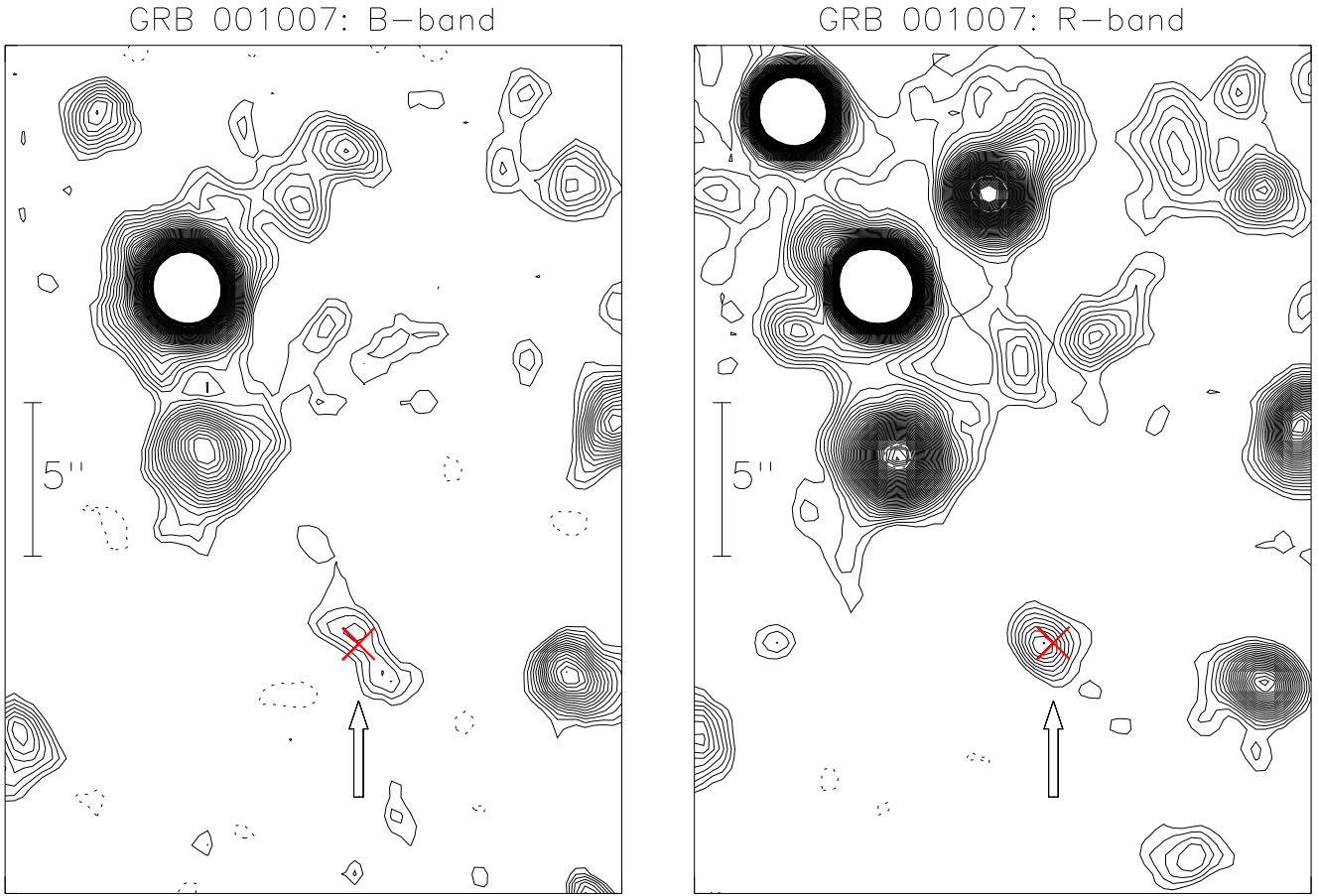}}
                 \caption{Contour plot displaying the host galaxy of \object{GRB 001007} and the position of its OA. {\bf Left panel:} it shows the coadded $B$ band image taken from 16.1203 UT to 16.1718 UT November 2000 at the 3.60ESO telescope, with still significant contribution from the OA. {\bf Right panel:} it shows the coadded $R$ band image taken from 13.3770 UT to 13.4092 UT, and from 14.2787 UT to 14.2999 UT, September 2001 at the 3.60ESO telescope, with negligible contribution from the OA. {\bf General:} Both sets of images revealed the presence of an extended and elongated object in the Northeast-Southwest direction (PA $\sim$ +45\degr). The position of the OA is indicated by a cross and, in both filtres, it is fully consistent with it. The $B$ band image suggested that the source was part of a complex object. Further VLT images confirmed (see Fig. \ref{trazado de contornos TMG}) that this extended source is in fact composed of at least two objects. The complexity of the object can not be visualised in the $R$ band image displayed in the right panel due to the non negligible contribution of the afterglow on 16.1203--16.1718 UT November 2000, 24.8\% of the total measured flux density. The position of the cross was calculated by transforming the image taken with the 1.54D (11.3192 UT--11.3673 UT October 2000, where the OA is clearly visible) to the frames of the 3.60ESO coadded images which are listed above. The error of the cross position (including frame alignment errors and internal astrometric errors) is $\sim$ 0.7\arcsec, smaller than its size (1\arcsec $\times$ 1\arcsec). See Fig. \ref{trazado de contornos TMG} for a more accurate view of the OA position. North is upwards and East leftwards.}
      \label{trazado de contornos 3,6}
  \end{figure*}

\subsection{Comparison of the light curve with predictions made by several models}

Many OAs exhibit a single power law decay index. Generally this index is $\alpha$ $\sim$ $-$1.3, a reasonable value for the spherical expansion of a relativistic blast wave in a constant density interstellar medium (\cite{meszaros-rees97}). Nonetheless there are OAs for which a break occurs within 2 days of the burst. This break has been seen and modelled as a signature of beaming. We note that collimated outflows in general result in faster decaying light curves than the spherically symmetric ones (\cite{huang00}).

We quantify a break as $\Delta\alpha$ = $\alpha_\mathrm{1} - \alpha_\mathrm{2}$, where $\alpha_\mathrm{1}$ is the decay index before the break and $\alpha_\mathrm{2}$ after the break. Six afterglow models have been considered in order to reproduce our values of $\alpha$ and $\beta$:

\emph{i}) An outflow arising in a jet expanding sideways in a homogeneous medium; then $\alpha_\mathrm{1}$ = 3($1 - p$)/4 and $\alpha_\mathrm{2}$ = $-p$ (\cite{rhoads99}). In this case, $p$ = 2.03 $\pm$ 0.11, and we would expect $\alpha_\mathrm{1}$ = $-$0.77 $\pm$ 0.08, and $\beta$ = $-$1.02 $\pm$ 0.06, what agrees with our data. This case would resemble \object{GRB 990510}, for which $\alpha_\mathrm{1}$ = $-$0.82 $\pm$ 0.02, $\alpha_\mathrm{2}$ = $-$2.18 $\pm$ 0.05 and a break occurred at $t_\mathrm{0}$ = 1.20 $\pm$ 0.08 days after the GRB (\cite{harrison99}).

\emph{ii}) A collimated outflow with a fixed opening angle in a homogeneous medium; then $\Delta\alpha$ = 0.75 (\cite{meszaros-rees99}). Here $\alpha_\mathrm{1}$ = 3($1 - p$)/4 changes to $\alpha_\mathrm{2}$ = $-$3$p$/4 in the optical range, considering an adiabatic outflow (\cite{panaitescu-meszaros99}). Following this model $p$ = 2.71 $\pm$ 0.15, $\alpha_\mathrm{1}$ = $-$1.28 $\pm$ 0.11, and $\beta$ = $-$1.35 $\pm$ 0.08, in agreement with the measured spectral index.

\emph{iii}) A collimated outflow with a fixed opening angle in an inhomogeneous medium (density gradient $\rho \propto r^{-s}$, with $s$ = 2 as expected in a stellar wind, \cite{panaitescu98}); then $\Delta\alpha$ = (3 $- s$)/(4 $- s$), changing from $\alpha_\mathrm{1}$ = $-$3($p - 1$)/4 $-$ $s$/(8 $-$ 2$s$) to $\alpha_\mathrm{2}$ = $-$3($p -$ 1)/4 $-$ (6 $- s$)/(8 $-$ 2$s$). For $s$ = 0 we recover \emph{ii}). In general, if the mean density distribution is not constant the light curve decays faster but, the break will be less pronounced. For the case of our $\alpha$ value, and assuming $s$ = 2, we derive $p$ = 2.37 $\pm$ 0.15, $\alpha_\mathrm{1}$ = $-$1.53 $\pm$ 0.11, and $\beta$ = $-$1.19 $\pm$ 0.08, also consistent with our value of $\beta$ = $-$1.24 $\pm$ 0.57. This case would be similar to \object{GRB 980519} where the decay index changed from $\alpha_\mathrm{1}$ = $-$1.73 $\pm$ 0.04 to $\alpha_\mathrm{2}$ = $-$2.22 $\pm$ 0.04 at $t_\mathrm{0}$ = 0.55 days after the gamma ray event (\cite{jaunsen01}).

The outcome of $\alpha_\mathrm{1}$ in the three jet models discussed above falls within the boundaries defined by the observations made to date: from $-$0.76 $\pm$ 0.01 in \object{GRB 990510} (\cite{harrison99}; \cite{stanek99}) to $-$1.73 $\pm$ 0.04 in \object{GRB 980519} (\cite{jaunsen01}).

\emph{iv}) The late evolution of highly relativistic jets of cannon balls emitted in supernova explosions. This model predicts a smooth knee having an after the break decay index $\alpha_\mathrm{2}$ = $-$2.1, and a spectral index $\beta$ = $-$1.1, fully consistent with our data (\cite{dado01}).

\emph{v}) A spherical adiabatic expansion with $\rho \propto r^{-s}$; we have a monotonic power law decay with $\alpha$ = $-$3($p~-$~1)/4~$-~s$/(8~$-$~2$s$). For $s$ = 2 (inhomogeneous medium due to a stellar wind) a value of $p$ = 3.04 $\pm$ 0.15 and $\beta$ = $-$1.52 $\pm$ 0.08 is expected. Although the derived value of $\beta$ is consistent with the measurements, the large value of $p$ is out of the range derived for the afterglows observed so far ($p$ ranges from 2 to 2.5). Thus, we consider this option less reconcilable with the data.

\emph{vi}) For $s$ = 0 (a spherical adiabatic expansion in an homogeneous medium) even more unrealistic values of $p$ are obtained ($p$ = 3.71). The inferred value of the spectral index ($\beta$ = $-p$/2 = -1.86) cannot be easily accommodated in the context of our measurements.

In view of the previous arguments we propose that the observed steep decay in the optical light curve may be due to a break which occurred before the optical observations started, $\sim$~3.5 days after the burst. The inferred values of $\Delta\alpha$, which can be explained in the context of \emph{i}), \emph{ii}), \emph{iii}) and \emph{iv}, are consistent with Fig. 3 from \cite{stanek01}. Furthermore, considering intrinsic extinction in the galaxy makes $\beta > -$1.24, therefore the spherical expansion models would be even more irreconcilable with our data. This conclusion agrees with the tentative knee suggested by the prompt 0.11LOTIS data (see Sect. \ref{curva de luz}.)

\subsection{The potential host galaxy}
     \label{presunta galaxia anfitriona}

Inspection of the $B$ band images taken in September 2001 with the 3.60ESO telescope suggested the presence of a faint object coincident with the position of the OA reported in Sect. \ref{resultados} (see cross of Fig. \ref{trazado de contornos 3,6}). Further $R$ band VLT imaging carried out $\sim 468$ days after the gamma ray event confirmed these evidences, detecting clearly the object previously imaged with the 3.60ESO. Its non-stellar profile (the angular extension of the source is $\sim$ 5\arcsec) and its consistency with the OA position makes it a strong candidate to be the host galaxy of \object{GRB 001007}.

\begin{figure}
      \resizebox{\hsize}{!}{\includegraphics{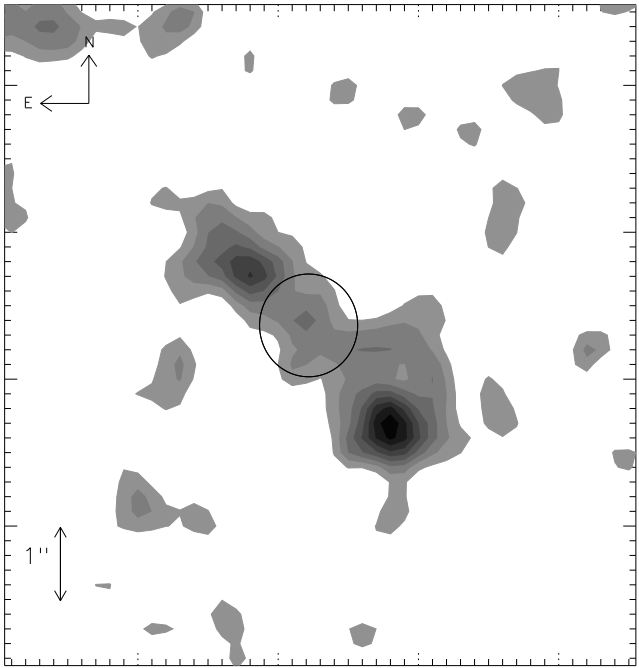}}
                 \caption{The figure shows the GRB 001007 field imaged with the VLT $\sim$ 468 days after the gamma ray event. The image corresponds to an integration time of 2\,400 s in the $R$ band. The circle indicates the position of the afterglow. Its radius corresponds to the astrometric and alignment errors (0.7\arcsec). As in Fig. \ref{trazado de contornos 3,6} the afterglow position has been derived using the image taken with the 1.54D (11.3192 UT--11.3673~UT October 2000) and transforming it to the VLT image frame. The image has been smoothed with a Gaussian filtre with a FWHM of 1 pixel (0.2\arcsec). As can be seen an extended object is consistent with the afterglow position. The contours scale linearly starting from 1$\sigma$ above the background. This smoothed image shows that the object is composed of at least two sources, 6$\sigma$ and 9$\sigma$ above the local background level. A possible third component (3$\sigma$ above the background) can be seen inside the astrometric circle.}
      \label{trazado de contornos TMG}
  \end{figure}

The excellent VLT observations (seeing of 0.65\arcsec) revealed the complex morphology of the host galaxy (see Fig. \ref{trazado de contornos TMG}). The host is composed of at least two objects, 6$\sigma$ and 9$\sigma$ above the local background level, respectively. The position of these two sources are: $\alpha_\mathrm{2000}$ = 4$^\mathrm{h}$5$^\mathrm{m}$54.19$^\mathrm{s}$, $\delta_\mathrm{2000}$ = $-$21\degr 53\arcmin 46.7\arcsec and $\alpha_\mathrm{2000}$ = 4$^\mathrm{h}$5$^\mathrm{m}$54.34$^\mathrm{s}$, $\delta_\mathrm{2000}$ = $-$21\degr 53\arcmin 44.6\arcsec~ (error 0.6\arcsec in both cases). Between these two sources there is a faint source (only 3$\sigma$ above the background level) located at $\alpha_\mathrm{2000}$ = 4$^\mathrm{h}$5$^\mathrm{m}$54.28$^\mathrm{s}$, $\delta_\mathrm{2000}$ = $-$21\degr 53\arcmin 45.3\arcsec. Only this third possible component is consistent with the 1$\sigma$ astrometric circle (see Fig. \ref{trazado de contornos TMG}). The other two bright components (1.7$\sigma$ and 2.7$\sigma$ from the OA position respectively) are consistent with the tail of the astrometric error, so a possible connection with the OA can not be discarded.

Circular aperture photometry (aperture diametre 6\arcsec) yielded $B$ = 25.10 $\pm$ 0.25 and $R$ = 24.84 $\pm$ 0.15 for the host galaxy complex. The undereddened (by Galactic extinction) fluxes imply a blue spectral distribution ($F_\nu \propto \nu^\beta$, with $\beta \sim $0.26) as expected in the UV rest frame for a star forming galaxy. In fact, 	 comparison between the VLT $R$ band and the 3.60ESO $B$ band images reveals that the closest system (the one to the North-East from the OA, see Fig. 5) to the OA position is the bluest one. From the fit we interpolate at the frequency corresponding to the $V$ band ($\nu_V$ = 5.45 $\times$ 10$^{14}$ Hz) and derive a $V$ band flux density of 0.42 $\mu$Jy. The inferred $V$ band flux density as well as the measured $BR$ band flux densities (0.44$\mu$Jy and 0.40$\mu$Jy, both corrected by Galactic extinction) were used in Sect. \ref{forma espectral} to subtract the contribution of the host to the total measured flux density.

\section{Conclusions}
  \label{conclusiones}

We presented observations of the OA associated with \object{GRB 001007} and its likely host galaxy; prompt images starting 6.14~hours after the event and late time images up to $t_\mathrm{0}$ + $\sim 468$~days. The $R$ band light curve is well fitted by a power law plus a constant brightness component due to the host galaxy. The decay indices in the $B$ and $R$ bands are $\alpha_B$ = $-$2.65 $\pm$ 0.73 and $\alpha_R$ = $-$2.03 $\pm$ 0.11. They are consistent with each other suggesting that the decay could be achromatic, so we assumed $\alpha$ = $\alpha_R$ throughout the paper. The $BVR$ band observations carried out at $\sim$ 11.15 UT October 2000 allowed us to determine a spectral index $\beta$ = $-$1.24 $\pm$ 0.57.

Several models were considered to explain our value of $\alpha$ and $\beta$. The existence of a jet is supported but we can not distinguish between the different jet geometries and density profiles of the GRB environment. An alternative explanation can be provided by the late evolution of highly relativistic jets of cannon balls emitted in supernova explosions. A spherical expansion in an inhomogeneous medium is marginally consistent with our data (because of poor sampling of the light curve and the SED). A spherical expansion in a homogeneous medium is inconsistent.

We proposed that the observed steep decay in the optical light curve may be due to a break which occurred before the OA was discovered, $\sim$3.5 days after the burst. A relativistic jet expansion (fireball or cannonball) could explain the fast decay as well as the proposed break. This suggestion agrees with the upper limits derived from our prompt optical observations. The extrapolation of the $R$ band light curve to epoch of the observations carried out by 0.11LOTIS, predicts $R$=14.63$\pm$0.34 and $V$ = 15.09 $\pm$ 0.36. These values are {\bf $1.2\sigma$} above the upper limit imposed by 0.11LOTIS, therefore consistent with the presence of a knee in the optical light curve.

The potential host galaxy is in a complex system with integrated magnitudes $B$ = 25.10 $\pm$ 0.25 and $R$ = 24.84 $\pm$ 0.15. Its overall spectral energy distribution is consistent with that of a star forming galaxy. The system shows a complex morphology, being composed of at least two components, with the bluest one being closer to the OA. Deep observations with the Hubble Space Telescope are required to study the nature of the system and the precise location of the OA within it, determining whether the possible knot seen at a 3$\sigma$ level at the OA position would be a star forming region associated to the bluest component or not.

\begin{acknowledgements}

J.M. Castro Cer\'on acknowledges the receipt of a FPI doctoral fellowship from Spain's Ministerio de Ciencia y Tecnolog\'\i a and the hospitality of the Danish Space Research Institute and the Instituto de Astrof\'\i sica de Andaluc\'\i a (IAA-CSIC) where part of this work was carried out. J. Gorosabel acknowledges the receipt of a Marie Curie Research Grant from the European Commission. K. Hurley is grateful for Ulysses support under JPL Contract 958\,056, and for IPN and NEAR support under NASA grants NAG5-3500 and NAG5-9503. This work was partially supported by the Danish Natural Science Research Council. The data presented here have been taken in part using ALFOSC, which is owned by the IAA-CSIC and operated at the Nordic Optical Telescope under agreement between the IAA-CSIC and the Astronomical Observatory of the NBIfAFG. Some of the observations presented in this paper were obtained under the ESO Large Programmes 165.H-0464 and 265.D-5742.

\end{acknowledgements}

\end{document}